\documentclass{svmult}
\usepackage[T1]{fontenc}
\usepackage[british]{babel}
\usepackage[utf8]{inputenc}
\usepackage[sf,SF]{subfigure}
\usepackage{mathptmx}
\usepackage{helvet}
\usepackage{courier}
\usepackage{makeidx}
\usepackage{multicol}
\usepackage{footmisc}
\usepackage{amsmath,amssymb}
\usepackage{tikz}
\usepackage{color}
\usepackage{url}
\usepackage{graphicx}
\usepackage{todonotes}

\usetikzlibrary{calc}
\usetikzlibrary{arrows}
\usetikzlibrary{positioning}
\usetikzlibrary{intersections}
\usetikzlibrary{decorations.pathreplacing}
\usetikzlibrary{shapes.geometric}

\graphicspath{{figures/}}

\makeatletter
\newcommand*{\dif}{\@ifnextchar^{\DIfF}{\DIfF^{}}}
\def\DIfF^#1{\mathop{\mathrm{\mathstrut d}}\nolimits^{#1}\gobblesp@ce}
\def\gobblesp@ce{\futurelet\diffarg\opsp@ce}
\def\opsp@ce{%
  \let\DiffSpace\!%
  \ifx\diffarg(%
    \let\DiffSpace\relax
  \else
    \ifx\diffarg[%
      \let\DiffSpace\relax
    \else
      \ifx\diffarg\{%
        \let\DiffSpace\relax
      \fi\fi\fi\DiffSpace}
\makeatother

\newcommand*{\dt}[1]{\ensuremath{\frac{\partial #1}{\partial t}}}
\newcommand*{\pd}[2]{\ensuremath{\frac{\partial #1}{\partial{#2}}}}

\newcommand*{\vct}[1]{\ensuremath{\vec{#1}}}

\renewcommand{\div}{\boldsymbol\nabla\cdot}
\newcommand{\grad}{\boldsymbol\nabla}

\newcommand*{\meter}{\textup{m}}
\newcommand*{\second}{\textup{s}}
\newcommand*{\pascal}{\textup{Pa}}
\newcommand*{\ifrac}[2]{#1/#2}

\newbox\bokstav
\newdimen\hoyde
\newcommand{\bgl}{{\hbox{$\left\lbrack\vbox to 8.5pt{}\right.\nOspace$}}}
\newcommand{\Bgl}{{\hbox{$\left\lbrack\vbox to 11.5pt{}\right.\nOspace$}}}
\newcommand{\bggl}{{\hbox{$\left\lbrack\vbox to 14.5pt{}\right.\nOspace$}}}
\newcommand{\Bggl}{{\hbox{$\left\lbrack\vbox to 17.5pt{}\right.\nOspace$}}}
\newcommand{\bgr}{{\hbox{$\left\rbrack\vbox to 8.5pt{}\right.\nOspace$}}}
\newcommand{\Bgr}{{\hbox{$\left\rbrack\vbox to 11.5pt{}\right.\nOspace$}}}
\newcommand{\bggr}{{\hbox{$\left\rbrack\vbox to 14.5pt{}\right.\nOspace$}}}
\newcommand{\Bggr}{{\hbox{$\left\rbrack\vbox to 17.5pt{}\right.\nOspace$}}}
\newcommand{\nOspace}{\nulldelimiterspace=0pt \mOth}
\newcommand{\mOth}{\mathsurround=0pt}
\newcommand{\Ljmp}{\mathopen{\lbrack\!\lbrack}}
\newcommand{\Rjmp}{\mathclose{\rbrack\!\rbrack}}
\newcommand{\bgLjmp}{\mathopen{\bgl\mskip-6mu\bgl}}
\newcommand{\bgRjmp}{\mathclose{\bgr\mskip-6mu\bgr}}
\newcommand{\BgLjmp}{\mathopen{\Bgl\!\!\Bgl}}
\newcommand{\BgRjmp}{\mathclose{\Bgr\!\!\Bgr}}
\newcommand{\bggLjmp}{\mathopen{\bggl\!\!\bggl}}
\newcommand{\bggRjmp}{\mathclose{\bggr\!\!\bggr}}
\newcommand{\BggLjmp}{\mathopen{\Bggl\!\!\Bggl}}
\newcommand{\BggRjmp}{\mathclose{\Bggr\!\!\Bggr}}
\newcommand{\jmp}[1]{%
\setbox\bokstav=\hbox{$ \left. #1\right. $}
\hoyde=\ht\bokstav
\advance\hoyde by \dp\bokstav%
\hbox{$
  \ifinner
    \ifdim\hoyde<10pt
      \Ljmp #1 \Rjmp%
    \else
      \ifdim\hoyde <11pt
        \Ljmp #1 \Rjmp%
      \else
        \ifdim\hoyde <14pt
          \bgLjmp #1 \bgRjmp%
        \else
          \ifdim\hoyde <20pt
            \BgLjmp #1 \BgRjmp%
          \else
            \bggLjmp #1 \bggRjmp%
          \fi
        \fi
      \fi
    \fi
  \else
    \ifdim\hoyde<8.5pt
      \Ljmp #1 \Rjmp%
    \else
      \ifdim\hoyde <11.5pt
        \bgLjmp #1 \bgRjmp%
      \else
        \ifdim\hoyde <14.5pt \Ch
          \BgLjmp #1 \BgRjmp%
        \else
          \ifdim\hoyde <17.5pt
            \bggLjmp #1 \bggRjmp%
          \else
            \BggLjmp #1 \BggRjmp%
          \fi
        \fi
      \fi
    \fi
  \fi
$}}

\begin{document}

\title*{Curvature calculations for the level-set method}
\author{Karl Yngve Lervåg and Åsmund Ervik}
\institute{Karl Yngve Lervåg, \email{karl.yngve@lervag.net}
\at
Norwegian University of Science and Technology,
Department of Energy and Process Engineering,
Kolbjørn Hejes veg 2, NO-7491 Trondheim, Norway.
\and Åsmund Ervik, \email{aaervik@gmail.com}
\at
SINTEF Energy Research,
Sem Sælands veg 11, NO-7465 Trondheim, Norway.\\
Norwegian University of Science and Technology,
Department of Physics,
Høgskoleringen 5, NO-7491 Trondheim, Norway.}
\maketitle

\abstract{The present work illustrates a difficulty with the level-set method
to accurately capture the curvature of interfaces in regions that are of equal
distance to two or more interfaces.  Such regions are characterized by kinks in
the level-set function where the derivative is discontinuous.  Thus the
standard discretization scheme is not suitable.  Three discretization schemes
are outlined that are shown to perform better than the standard discretization
on two selected test cases.}

\section{Introduction}
This article addresses the calculation of interface curvature with the
level-set method.  In the level-set method, the normal vector and the curvature
of an interface can be calculated directly from the level-set function.  These
calculations are usually done with standard finite-difference methods,
typically the second-order central difference scheme (CD-2)
\cite{Osher88,Sethian03,Kang00}.

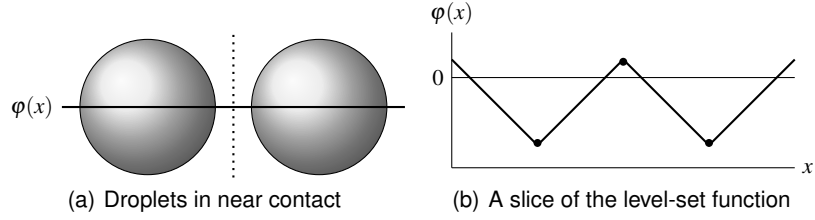
\begin{figure}[btp]
 \centering
 \subfigure[Droplets in near contact]{
 \begin{tikzpicture}
 [
 scale=0.6,
 drop/.style={shading=ball, ball color=black!10},
 ]

 \shadedraw[drop] (-1.9,0) circle (1.5);
 \shadedraw[drop] ( 1.9,0) circle (1.5);
 \draw[thick] (-3.8,0.0) node[left] {$\varphi(x)$} -- (3.8,0.0);
 \draw[thick, dotted] (0,1.6) -- (0,-1.6);
 \end{tikzpicture}
 \label{fig:1a}}
 \subfigure[A slice of the level-set function]{
 \begin{tikzpicture}[scale=0.6]
 \draw (-3.8,2.5) node[above] {$\varphi(x)$} -- (-3.8,-0.5);
 \draw (-3.8,-0.5) -- ( 3.8,-0.5) node[right] {$x$};
 \draw[thin,black] (-3.8,1.5) node[left] {$0$} -- (3.8,1.5);
 \coordinate (a) at (-3.8,1.9);
 \coordinate (b) at (-1.9,0);
 \coordinate (c) at ( 0.0,1.9);
 \coordinate (d) at ( 1.9,0);
 \coordinate (e) at ( 3.8,1.9);
 \draw[thick] (a) -- (b) -- (c) -- (d) -- (e);

 \fill ( 0.0,1.85) circle (2.5pt);
 \fill (-1.9,0.05) circle (2.5pt);
 \fill ( 1.9,0.05) circle (2.5pt);
 \end{tikzpicture}
 \label{fig:1b}}
 \caption{(a) Two droplets in near contact.  The dotted line marks a region
 where the derivative of the level-set function is not defined. (b)
 A one-dimensional slice of the level-set function $\varphi(x)$.  The dots mark
 points where the derivative of $\varphi(x)$ is not defined.}
 \label{fig:cdroplets}
\end{figure}

A problem with these calculations may arise when the level-set function is
defined to be a signed-distance function.  The signed-distance function is in
general not smooth, as can be seen in Figure~\ref{fig:cdroplets}.  Here the
derivative of the level-set function will be discontinuous at the regions that
are of equal distance to more than one interface.  When two droplets as in
Figure~\ref{fig:cdroplets} are in near contact, such discontinuities, or kinks, may
lead to significant errors when calculating the interface geometries with
standard finite difference methods.

\section{Governing equations}
\subsection{Navier-Stokes equations for two-phase flow}
Consider a domain $\Omega=\Omega^+\cup\,\Omega^-$, where $\Omega^+$ and
$\Omega^-$ denote regions occupied by two respective phases, divided by an
interface $\Gamma=\delta\Omega^+\cap\delta\Omega^-$.  The governing equations
for incompressible and immiscible two-phase flow in the domain $\Omega$ with an
interface force on the interface $\Gamma$ are
\begin{align}
  \label{eq:masseq}
  \div\vct u &= 0, \\
  \label{eq:momeq}
  \rho \left( \dt{\vct u} + \vct u\cdot\grad\vct u\right) &= -\grad
  p + \div(\mu\grad\vct u) + \rho\vct f_b + \int_{\Gamma} \sigma\kappa\vct
  n\,\delta(\vct x-\vct x_I(s))\dif s.
\end{align}
Here $\vct u$ is the velocity vector, $p$ is the pressure, $\vct f_b$ is the
specific body force, $\sigma$ is the coefficient of surface tension, $\kappa$
is the curvature, $\vct n$ is the normal unit vector which points into
$\Omega^+$, $\delta$~ is the Dirac delta function, $\vct x_I(s)$ is
a parametrization of the interface, $\rho$ is the density and $\mu$ is the
viscosity.

It is assumed that the density and viscosity are constant in each phase, but
may be discontinuous across the interface.  The jump conditions across the
interface are
\begin{align}
  \jmp{\vct u} &= 0, \\
  \jmp{p} &= 2\jmp{\mu}\vct n\cdot\grad\vct u\cdot\vct n + \sigma\kappa, \\
  \jmp{\mu\grad\vct u} &= \jmp{\mu} \big(
      (\vct n\cdot\grad\vct u\cdot\vct n)\vct n\vct n +
      (\vct n\cdot\grad\vct u\cdot\vct t)\vct n\vct t +
      (\vct n\cdot\grad\vct u\cdot\vct t)\vct t\vct n +
      (\vct t\cdot\grad\vct u\cdot\vct t)\vct t\vct t
    \big),
\end{align}
where $\vct t$ is the tangent vector along the interface and $\jmp{\cdot}$
denotes the jump across an interface, that is $\jmp{\mu} \equiv \mu^+ - \mu^-$.
Note that $\grad\vct u$ and (e.g.) $\vct n\vct t$ are rank-2 tensors.  See
\cite{Kang00,Hansen05} for more details and a derivation of the interface
conditions.

\subsection{Level-set method}
The interface is captured with the zero level set of the level-set function
$\varphi(\vct x,t)$, which is prescribed as a signed-distance function.  It is
updated by solving an advection equation for $\varphi$,
\begin{equation}
  \label{eq:lseq}
  \dt{\varphi} + \vct{\hat u}\cdot\grad\varphi = 0,
\end{equation}
where $\vct{\hat u}$ is the velocity at the interface, extended to the entire
domain by solving
\begin{equation}
  \label{eq:lsvelext}
  \pd{\vct{\hat u}}{\tau} + S(\varphi)\vct n\cdot\grad\vct{\hat u} = 0,
  \quad \vct{\hat u}_{\tau=0} = \vct u,
\end{equation}
to steady state, cf.\ \cite{Zhao96}.  Here $\tau$ is a pseudo-time and
$S(\varphi) = \ifrac{\varphi}{(\varphi^2+2\Delta x^2)^{1/2}}$ is a smeared sign
function which is equal to zero at the interface.

When \eqref{eq:lseq} is solved numerically, the level-set function loses its
signed-distance property due to numerical dissipation.  The level-set function
is therefore reinitialized regularly by solving
\begin{equation}
  \label{eq:lsreinit}
  \begin{split}
    \pd{\varphi}{\tau} + S(\varphi_0)(|\grad\varphi|-1) &= 0, \\
    \varphi(\vct x,0) &= \varphi_0(\vct x),
  \end{split}
\end{equation}
to steady state as proposed in \cite{Sussman94}, where $\varphi_0$ is the
level-set function that needs to be reinitialized.

Normal vectors and curvatures can be readily calculated from the level-set
function as
\begin{equation}
  \label{eq:norm-curv}
  \vct n = \frac{\grad\varphi}{|\grad\varphi|} \quad \text{and} \quad
  \kappa = \div\left(\frac{\grad\varphi}{|\grad\varphi|}\right).
\end{equation}

\section{Numerical methods}
The Navier-Stokes equations \eqref{eq:masseq} and \eqref{eq:momeq} are solved using
a projection method on a staggered grid as described in
\cite[Chapter~5.1.1]{Hansen05}.  The spatial terms are discretized with CD-2,
except for the convective terms which are discretized by a fifth-order WENO
scheme.  A third-order strong stability-preserving Runge-Kutta (SSP RK) method
is used for the momentum equation \eqref{eq:momeq}, and
a second-order SSP-RK method is used for the level-set equations
\eqref{eq:lseq} to \eqref{eq:lsreinit} \cite{Gottlieb01}.

The interface conditions are treated in a sharp fashion with the Ghost-Fluid
Method (GFM), which incorporates the discontinuities into the discretization
stencils by altering the stencils close to the interfaces, cf.\ 
\cite{Fedkiw99,Kang00,Liu00}.  When using the GFM, the curvature is linearly
interpolated from the grid points to the interface before it is used in the
discretization stencils for the flow equations unless otherwise stated.

\section{Curvature discretizations}
The normal vector and the curvature \eqref{eq:norm-curv} are typically
discretized with the CD-2 at the grid points, cf.\
\cite{Kang00,Sethian03,Xu06}.  A problem with this is that CD-2 will not
converge across kinks, and it may therefore introduce potentially large errors.
The errors in the curvature will lead to erroneous pressure jumps at the
interfaces, and the errors in the normal vector affect both the discretized
interface conditions and the extrapolated velocity \eqref{eq:lsvelext} which is
used in the advection equation \eqref{eq:lseq}.

A direction difference scheme is presented in \cite{Macklin05} which uses
a combination of one-sided and central difference schemes to ensure that the
differences never cross kinks.  The same scheme is used in the present work to
calculate the normal vector.  The idea is choose which difference scheme to use
based on the values of a quality function,
\begin{equation}
  Q(\vct x) = \left| 1 - |\grad\varphi(\vct x)|\right|.
\end{equation}
The quality function is itself calculated with central differences.  It
effectively detects the regions where the level-set function differs from the
signed-distance function.  Let $Q_{i,j} = Q(\vct x_{i,j})$ and $\eta>0$, then
$Q_{i,j}>\eta$ can be used to detect kinks.  The parameter $\eta$ is tuned such
that the quality function will detect all the kinks.  The value $\eta=0.1$ is
used in the present work.

In the following, three different improved discretization schemes for the
curvature are outlined.  Note that the first two schemes use the quality
function to detect when the improved schemes should be used in favor of CD-2.
Also note that the curvature is only calculated at grid points in a narrow band
along the interface.  At the points where it is not calculated, it is set to
zero.

\emph{Macklin and Lowengrub's method} (MLM) was presented in
\cite{Macklin06,Macklin08}.  With this method, the interface is parametrized
with a second-order least-squares polynomial.  The curvature is then calculated
directly from the parametrization at the desired position on the interface.

To enable easy comparison with the other methods, the estimated curvature
values are extrapolated from the interface to the adjacent grid points.

\emph{Lervåg's method} (LM) was presented in \cite{Lervaag11-a} and is based on
MLM, specifically \cite{Macklin06}.  The curve parametrization is used to
create a local level-set function from which the curvature is calculated on
the grid points using CD-2.

The main difference from MLM is that the curvature is calculated at the grid
nodes and then interpolated to the interface afterwards.  This is argued as
a slight simplification of MLM, although an important consequence is that it
becomes more important to have an accurate representation of the interface.
Instead of using a least-squares parametrization, LM uses monotone cubic
Hermite splines.

\emph{Salac and Lu's method} (SLM) was presented in \cite{Salac08} and is
a different approach than MLM and LM.  Consider the 2D case of two circles in
near contact, see Figure~\eqref{fig:slm}.  SLM reconstructs two independent level-set
functions $\phi_{1}$ and $\phi_{2}$ for the two circles.  The reconstructed
functions are then used to calculate the curvature.  Since the two
reconstructed cones have no kinks, the curvature can be calculated with CD-2.
For points close to both circles, a weighted average of the curvature from
$\phi_{1}$ and from $\phi_{2}$ is stored.  For points close to only one circle,
the appropriate curvature is stored.  The weighted average is $\kappa
= (\kappa_{1}\phi_{2} + \kappa_{2}\phi_{1})/(\phi_{1} + \phi_{2})$, where the
subscripts refer to values calculated on the reconstructed level-set functions.
This weighting will prefer $\kappa_{1}$ when closest to circle $1$, and vice
versa.
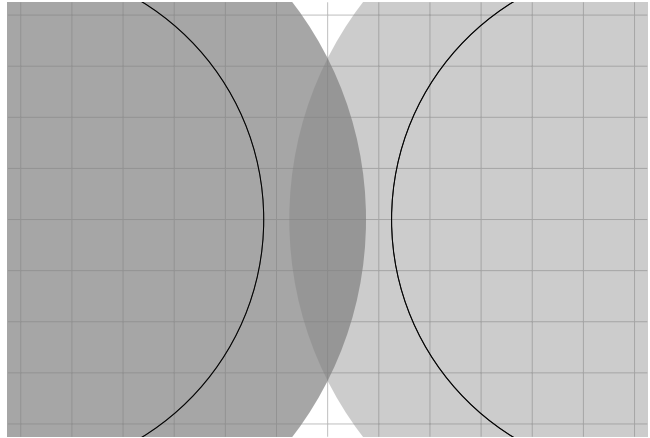
\begin{figure}[tbp]
  \centering
  \begin{tikzpicture} [
    scale=0.68,
    axes/.style={gray!50,very thin},
    local axes/.style={red,very thin,dashed},
    point/.style={black},
    interface/.style={very thin,black},
    approx/.style={thin,dashed,red},
    curve/.style={thick},
    ridge/.style={thin,red},
    >=latex,
    ]

    \coordinate (grid start) at (-2.25, -0.25);
    \coordinate (grid end)   at (10.25,  8.25);
    \coordinate (center1) at (-2.25, 4.00);
    \coordinate (center2) at (10.25, 4.00);
    \coordinate (pI) at (2.34, 6.00);
    \coordinate (p0) at (3.00, 6.00);
    \coordinate (p0 alt) at (2.00, 6.00);
    \clip (grid start) rectangle (grid end);

    \draw[axes]      (grid start) grid (grid end);
    \draw[interface] (center1) circle (5cm);
    \draw[interface] (center2) circle (5cm);

    \fill[fill opacity=0.7,gray] (center1) circle(7.00cm);
    \fill[fill opacity=0.4,gray] (center2) circle(7.00cm);
    \draw[interface] (center1) circle (5cm);
    \draw[interface] (center2) circle (5cm);
  \end{tikzpicture}
  \caption{Simple sketch of how SLM works.  The two circles are represented by
  separate level-set functions.}
  \label{fig:slm}
\end{figure}

\section{Comparison of the discretization schemes}
\subsection{A static disc above a rectangle}
Consider a disc of radius $r$ positioned at a distance $h$ above a rectangle,
see Figure~\ref{fig:circleandline}.  In this case, only the level-set function and
the geometrical quantities are considered.  None of the governing equations
\eqref{eq:masseq}, \eqref{eq:momeq} and \eqref{eq:lseq} to \eqref{eq:lsreinit} are solved.
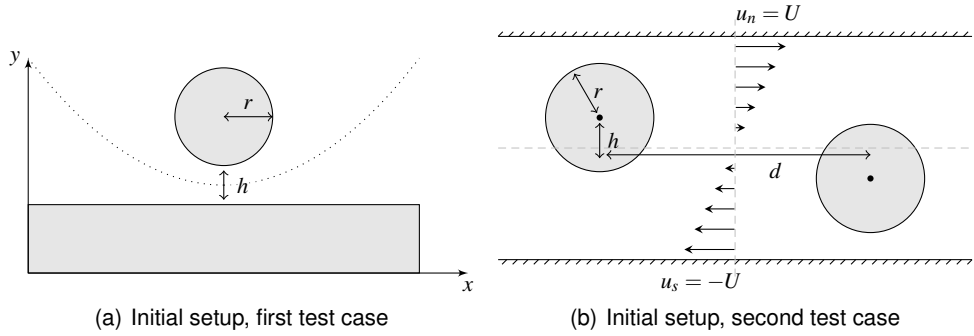
\begin{figure}[tbp]
  \centering
  \noindent\makebox[\textwidth]{ 
    \subfigure[Initial setup, first test case]{
      \begin{tikzpicture}
        [
        scale=1.3,
        hnode/.style = {fill=white,fill opacity=0.5,text opacity=1.0,right=2pt},
        ]
        \draw[fill=gray!20] (0,0.9) circle (0.5);
        \draw[<->] (0,0.9) -- node[above] {$r$} (0.5,0.9);

        \draw[dotted] (0,0.2) parabola (2,1.5);
        \draw[dotted] (0,0.2) parabola (-2,1.5);

        \draw[fill=gray!20] (-2,-0.7) rectangle (2,0);
        \draw[<->] (0,0.05) -- node[hnode] {$h$} (0,0.35);

        \draw[>=latex',->] (-2,-0.7) -- ( 2.5,-0.7) node[below] {$x$};
        \draw[>=latex',->] (-2,-0.7) -- (-2.0, 1.5) node[left]  {$y$};
      \end{tikzpicture}
      \label{fig:circleandline}
    } %
    \subfigure[Initial setup, second test case]{
      \begin{tikzpicture}
        [
        scale=0.9,
        wall/.style={
          decoration={border,angle=45,segment length=4},
          postaction={decorate,draw}},
        drop/.style={fill=gray!20},
        axis/.style={very thin, gray!50, densely dashed},
        u/.style={->, >=stealth,},
        ]

        \coordinate (c1) at (-2, 0.45);
        \coordinate (c2) at ( 2,-0.45);
        \draw[drop] (c1) circle(0.8);
        \draw[drop] (c2) circle(0.8);
        \fill (c1) circle(1.3pt);
        \fill (c2) circle(1.3pt);
        \draw[<->,rotate=30] (c1) +(0,0.08) -- node[right]{$r$} +(0.0,0.75);
        \draw[<->] (c1) +(0.0,-0.08) -- node[right]{$h$} +(0.0,-0.60);
        \draw[<->] (c1) +(0.1,-0.55) -- ($(c1) + (4.0,-0.55)$);
        \node[below] at (0.6,-0.1) {$d$};

        \draw[wall] (-3.5, 1.65) -- ( 3.5, 1.65);
        \node[above=2pt] at ( 0.5, 1.65) {$u_n=U$};
        \draw[wall] ( 3.5,-1.65) -- (-3.5,-1.65);
        \node[below=2pt] at (-0.5,-1.65) {$u_s=-U$};
        \draw[axis] (-3.5, 0.0) -- ( 3.5, 0.0);
        \draw[axis] ( 0.0,-1.9) -- ( 0.0, 1.9);
        \foreach \y/\x in {0.3/0.15,0.6/0.3,0.9/0.45,1.2/0.6,1.5/0.75} {
          \draw[u] ( 0.01, \y) -- ( \x, \y);
          \draw[u] (-0.01,-\y) -- (-\x,-\y); }
      \end{tikzpicture}
      \label{fig:drop-in-shear}
    }
  }
  \caption{Initial setup for the circle and rectangle test, (a), and for the
  drop collision in shear flow test, (b). In (a), the dotted line
  depicts the kink location, and there is no flow. In (b) the flow is indicated
  by the velocity profile.}
\end{figure}

The parameters used for this case are $r=0.25\ \meter$ and $h = \Delta x$.  The
domain is $1.5\ \meter\times 1.5\meter$, and the rectangle height is $0.75\
\meter$.  The grid size is $101\times 101$.

Figure~\ref{fig:simple-curvatures} shows a comparison of the calculated curvatures.
The figure shows that CD-2 leads to large errors in the calculated curvatures
in the areas that are close to two interfaces.  In particular note that the
sign of the curvature becomes wrong.  The analytic curvature for this case is
$\kappa = -1/r = -4$, and the curvature spikes seen for the standard
discretization is of the order of $|\kappa| \sim \frac{1}{\Delta x} \simeq
67.3$.  All of the improved methods give much better estimates of the
curvature, as expected.
\begin{figure}[tbp]
  \centering
  \noindent\makebox[\textwidth]{
  \subfigure[CD-2]{
    \includegraphics[width=0.30\textwidth]{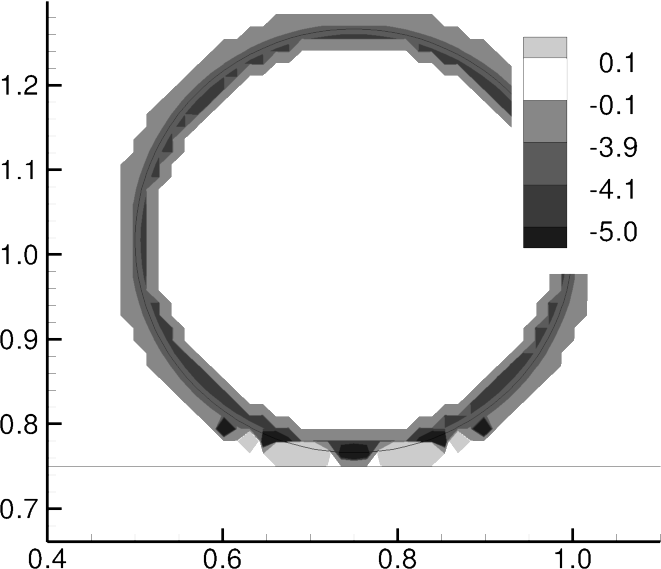}}
  \subfigure[MLM]{
    \includegraphics[width=0.30\textwidth]{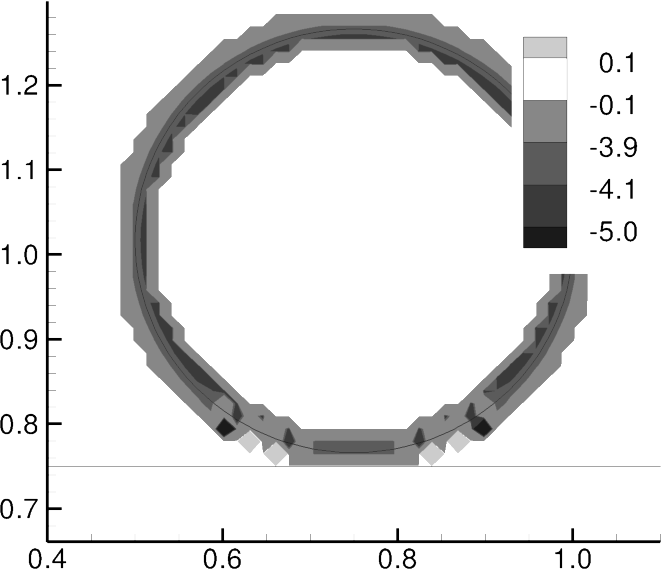}}
  \subfigure[LM]{
    \includegraphics[width=0.30\textwidth]{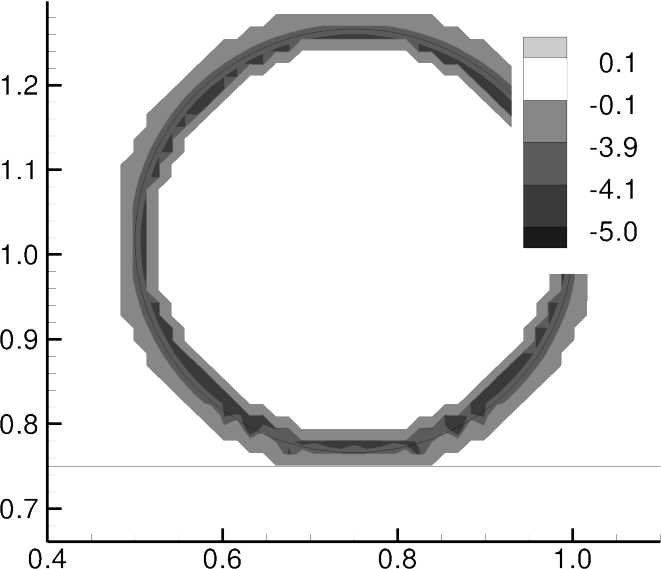}}
  \subfigure[SLM]{
    \includegraphics[width=0.30\textwidth]{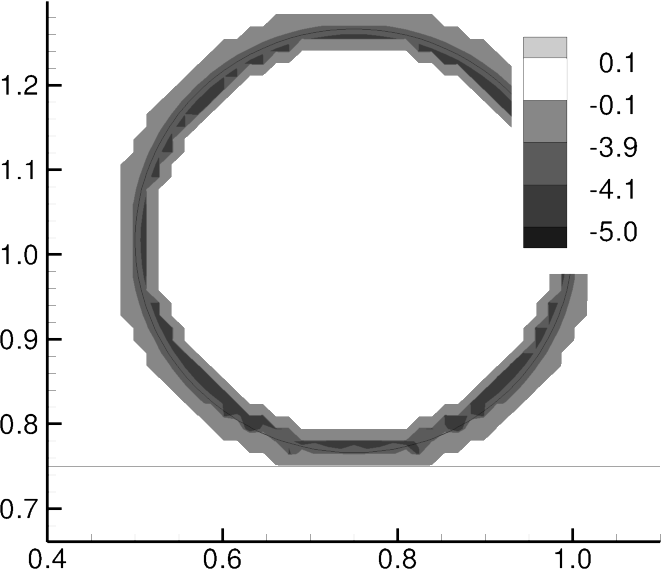}}
  }
  \caption{A comparison of curvature calculations between standard
  discretization and the improved method.  The standard discretization leads to
  large errors in the curvatures in areas that are close to two interfaces.}
  \label{fig:simple-curvatures}
\end{figure}

\subsection{Drop collision in shear flow}
Now consider two drops in a shear flow as depicted in Figure~\ref{fig:drop-in-shear}.
Both drops have radius $r$ and are initially placed a distance
$d=5r$ apart in the shear flow, where the flow velocity changes linearly from
$u_s=-U<0$ at the bottom wall to $u_n=U$ at the top wall.  The computational
domain is $12r\times8r$, and the grid size is $241\times 161$.  The density and
viscosity differences of the two phases are zero.

The shear flow is defined by the Reynolds number and the Capillary number,
\begin{equation}
  Re = \frac{\rho Ur}{\mu}
  \qquad\text{and}\qquad
  Ca = \frac{\mu U}{\sigma}.
\end{equation}
The following results were obtained with $r = 0.5\ \meter$, $h = 0.84r = 0.42\
\meter$, $Re = 10$ and $Ca=0.025$.

Figure~\ref{fig:ycf84} shows a comparison of the interface evolution and the
curvature between the different discretization schemes.  The first column shows
the results with the CD-2.  The next three columns show the results with the
three improved schemes respetively.  The kinks between the drops lead to
curvature spikes with CD-2, whereas the improved discretizations calculate the
curvature along the kink in a much more reliable manner.  LM and SLM give very
similar results.  This is most likely due to the fact that both these methods
calculate the curvature at the grid points and then interpolate, resulting in
very similar algorithms as long as the curvature calculations are accurate.
MLM on the other hand removes the interpolation step and calculates the
curvature directly on the interface.  Note that the difference is mainly that
the MLM results in slightly earlier coalescence in the given case.

The curvature spikes in obtained with CD-2 are seen to prevent coalescence.
This is due to the effect they have on the pressure field as displayed in
Figure~\ref{fig:ycf84-pressure}.  Here it is shown that the errors in the curvature
with CD-2 lead to an erroneous pressure field between the drops.  The
distortion of the pressure in the thin-film region leads to a flow into the
film region that suppresses coalescence.  The corresponding result with LM
shows that when the pressure is not distorted, it leads to a flow directed out
of the thin-film region.
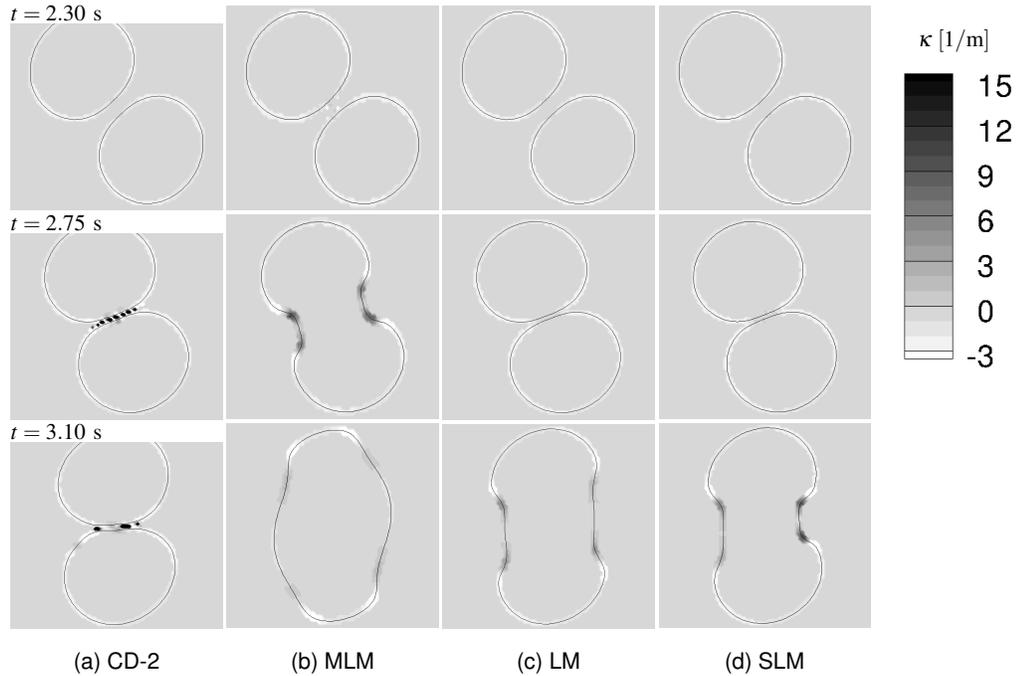
\begin{figure}[tbp]
  \centering
  \begin{tikzpicture}
    [
    time/.style={fill=white,text width=0.24\textwidth,inner sep=1pt},
    caption/.style={font=\sffamily},
    ]

    \node (standard1) at (0,0)
      {\includegraphics[width=0.24\textwidth]{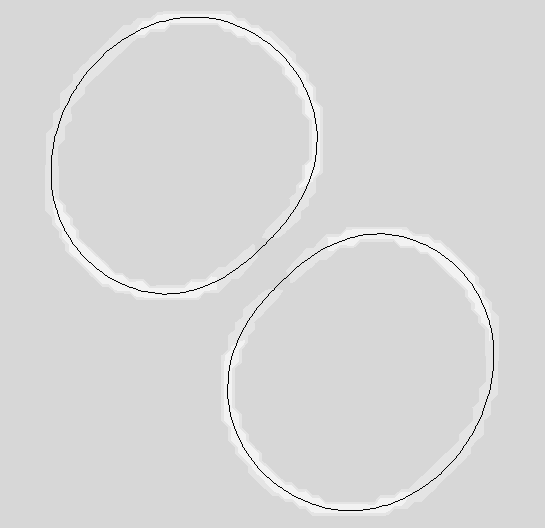}};
    \node (standard2) [below=-0.15cm of standard1]
      {\includegraphics[width=0.24\textwidth]{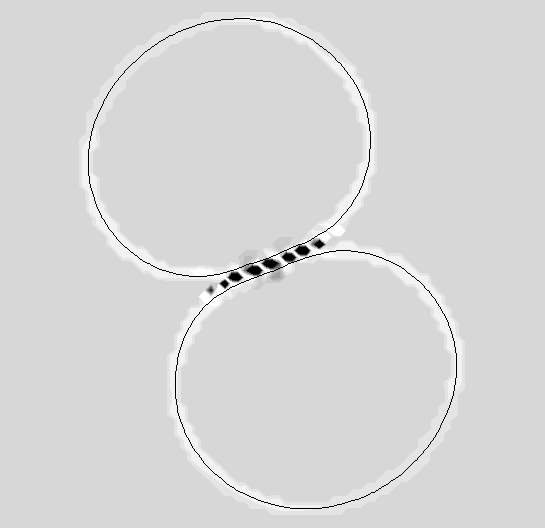}};
    \node (standard3) [below=-0.15cm of standard2]
      {\includegraphics[width=0.24\textwidth]{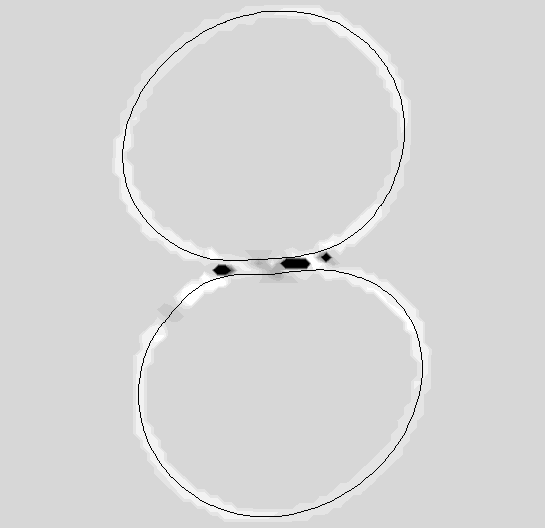}};
    \node (macklin1) [right=-0.15cm of standard1]
      {\includegraphics[width=0.24\textwidth]{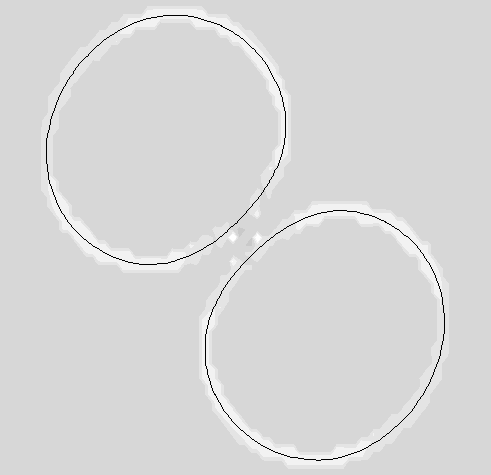}};
    \node (macklin2) [below=-0.15cm of macklin1]
      {\includegraphics[width=0.24\textwidth]{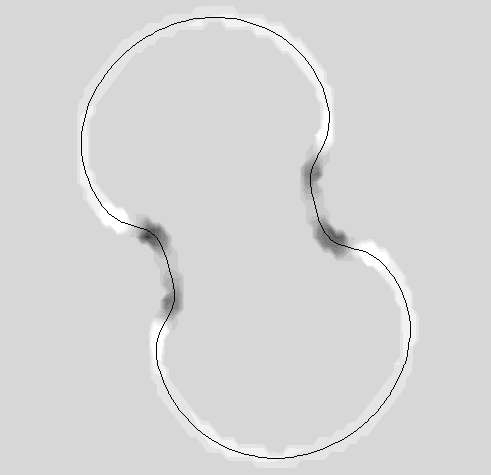}};
    \node (macklin3) [below=-0.15cm of macklin2]
      {\includegraphics[width=0.24\textwidth]{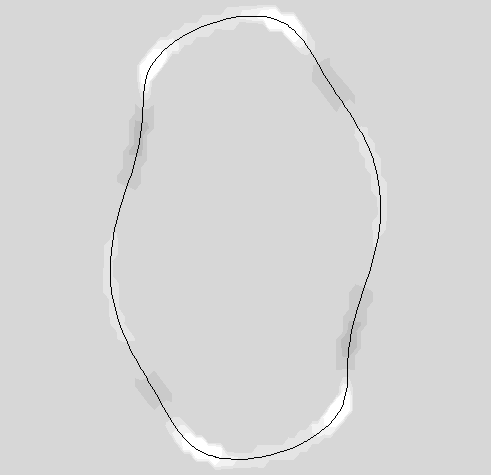}};
    \node (lervag1) [right=-0.15cm of macklin1]
      {\includegraphics[width=0.24\textwidth]{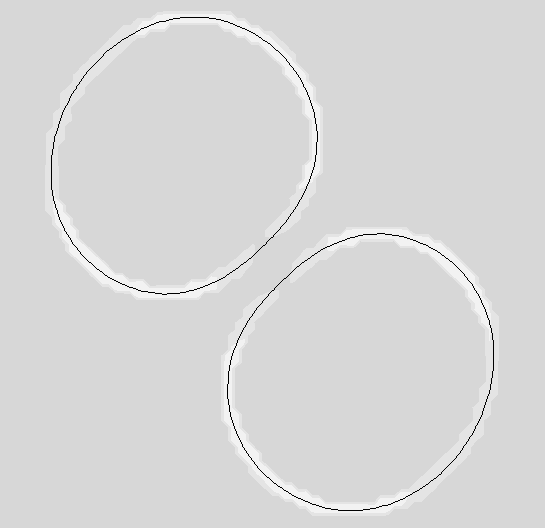}};
    \node (lervag2) [below=-0.15cm of lervag1]
      {\includegraphics[width=0.24\textwidth]{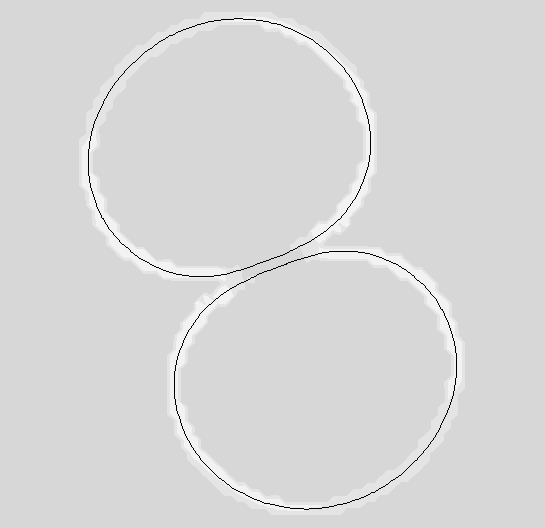}};
    \node (lervag3) [below=-0.15cm of lervag2]
      {\includegraphics[width=0.24\textwidth]{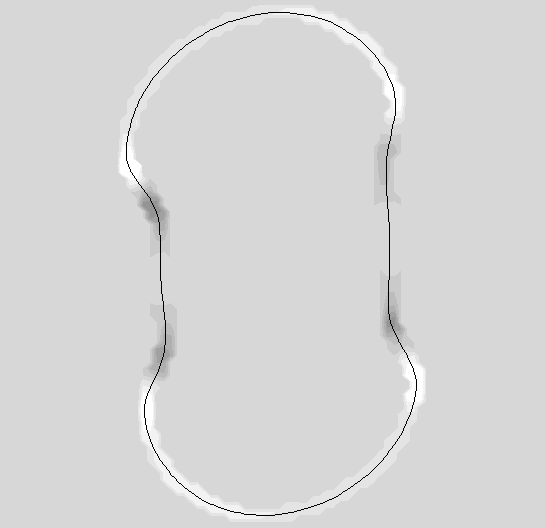}};
    \node (salac1) [right=-0.15cm of lervag1]
      {\includegraphics[width=0.24\textwidth]{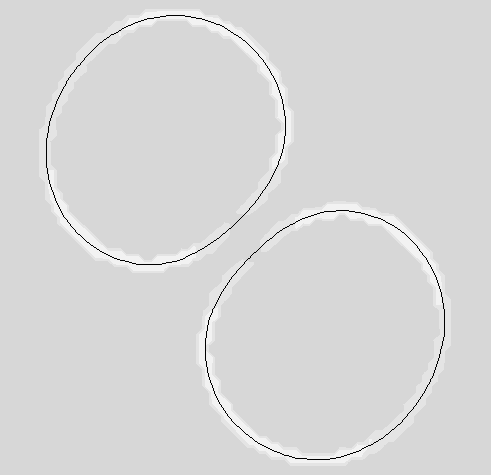}};
    \node (salac2) [below=-0.15cm of salac1]
      {\includegraphics[width=0.24\textwidth]{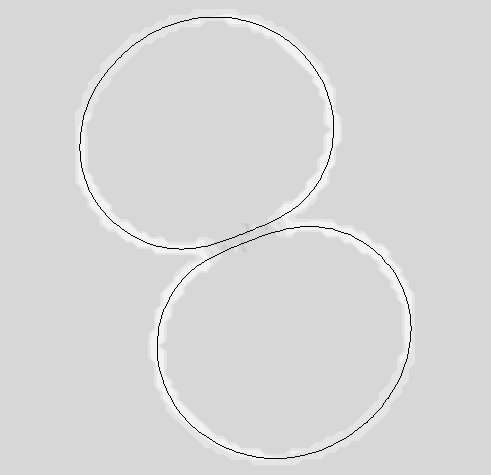}};
    \node (salac3) [below=-0.15cm of salac2]
      {\includegraphics[width=0.24\textwidth]{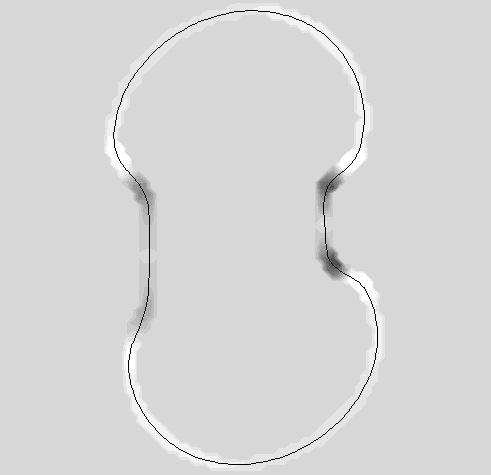}};
    \node[time,above=-10pt of standard1] {$t=2.30\ \second$};
    \node[time,above=-10pt of standard2] {$t=2.75\ \second$};
    \node[time,above=-10pt of standard3] {$t=3.10\ \second$};

    \node[caption,below=3pt of standard3] {(a) CD-2};
    \node[caption,below=3pt of macklin3]  {(b) MLM};
    \node[caption,below=3pt of lervag3]   {(c) LM};
    \node[caption,below=3pt of salac3]    {(d) SLM};

    \node (legend) [right=0.0cm of salac1.south east]
      {\includegraphics[width=1.8cm]{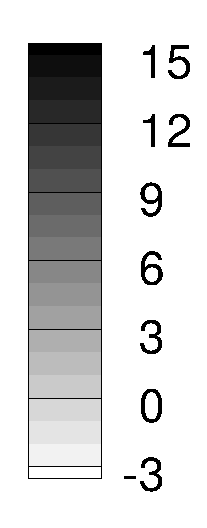}};
    \node [above=-0.3cm of legend] {$\kappa\ [1/\meter]$};

  \end{tikzpicture}
  \caption{A comparison between the different discretization schemes of the
  interface evolution and the curvature $\kappa$ of drop collision in shear
  flow.}
  \label{fig:ycf84}
\end{figure}
\begin{figure}[tbp]
  \centering
  \subfigure[CD-2]{
    \includegraphics[width=0.40\textwidth]{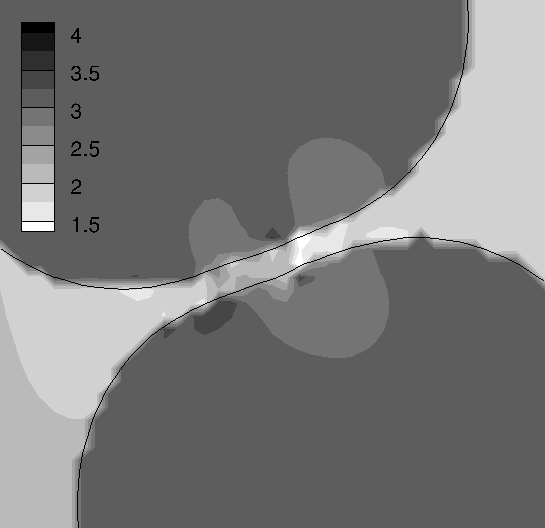}
    \label{fig:ycf84-pressure-nolc}
    }
  \subfigure[LM]{
  \includegraphics[width=0.40\textwidth]{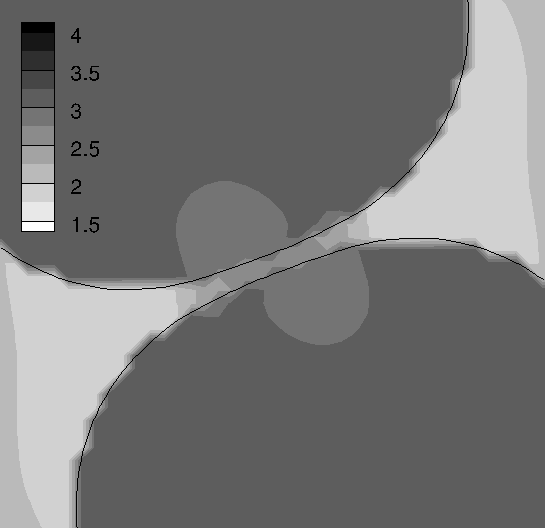}
    \label{fig:ycf84-pressure-lc}
    } \caption{Comparison of the pressure field in the thin film between the
    droplets at $t=2.75\ \second$.  The contour legends indicate the pressure
    in \pascal.}
  \label{fig:ycf84-pressure}
\end{figure}

\section{Conclusions}
Three discretization schemes have been implemented to accurately calculate the
curvature in regions close to kinks in the level-set function.  It has been
demonstrated in two test cases that the standard second-order central
difference scheme (CD-2) leads to relatively severe errors across the kinks.
Macklin and Lowengrub's method (MLM), Lervåg's method (LM), and Salac and Lu's
method (SLM) all give better results.  In the second test case where two
droplets are put in a shear flow, CD-2 gives a qualitatively different result
than all the three improved schemes due to an erroneous pressure field in the
thin film region.

\section*{Acknowledgements}
The authors acknowledge Bernhard Müller (Norwegian University of Science and
Technology) and Svend Tollak Munkejord (SINTEF Energy Research) for valuable
feedback on the manuscript.

This work was financed through the Enabling Low-Emission LNG Systems project,
and the authors acknowledge the contributions of GDF SUEZ, Statoil and the
Petromaks programme of the Research Council of Norway (193062/S60).

\bibliographystyle{spmpsci}

\end{document}